\documentclass[10pt,twocolumn]{article}
\usepackage{isqcmc2025} 

\usepackage{graphicx,url}
\usepackage{float}
\usepackage{enumerate}
\usepackage{amsmath}

\usepackage[latin1]{inputenc}

\usepackage[blind]{anonymize} 
\usepackage{listings}

\lstdefinelanguage{OPENQASM}{
morekeywords={q, c}, 
morekeywords=[2]{OPENQASM, include}, 
emph={h,cx,qreg,creg,->}, 
morekeywords=[3]{measure}, 
morekeywords=[4]{barrier}, 
sensitive=true,
morecomment=[l]{//}, 
morestring=[b]",
literate={->}{{\textbf{\color{codeemph2}{$\to$}}}}1 
}

\title{Parallel DNA Sequence Alignment on High-Performance Systems with CUDA and MPI}

\author{Linus Zwaka\thanks{Undergraduate student at RPI}
}
\address{Rensselaer Polytechnic Institute (RPI), Troy, NY
         \email{zwakal@rpi.edu}
}

\begin{document}

\maketitle

\begin{abstract}
Sequence alignment is a cornerstone of bioinformatics, widely used to identify similarities between DNA, RNA, and protein sequences and studying evolutionary relationships and functional properties. The Needleman-Wunsch algorithm remains a robust and accurate method for global sequence alignment. However, its computational complexity, 
O(mn), poses significant challenges when processing large-scale datasets or performing multiple sequence alignments. To address these limitations,  a hybrid implementation of the Needleman-Wunsch algorithm that leverages CUDA for parallel execution on GPUs and MPI for distributed computation across multiple nodes on a supercomputer is proposed. CUDA efficiently offloads computationally intensive tasks to GPU cores, while MPI enables communication and workload distribution across nodes to handle large-scale alignments.

This work details the implementation and performance evaluation of the Needleman-Wunsch algorithm in a massively parallel computing environment. Experimental results demonstrate significant acceleration of the alignment process compared to traditional CPU-based implementations, particularly for large input sizes and multiple sequence alignments. In summary, the combination of CUDA and MPI effectively overcomes the computational bottlenecks inherent to the Needleman-Wunsch algorithm without requiring substantial modifications to the underlying algorithm, highlighting the potential of high-performance computing in advancing sequence alignment workflows.

\end{abstract}

\section{Introduction}\label{sec:gen}

Sequence alignment is a cornerstone of bioinformatics, widely used to identify similarities between DNA, RNA, and protein sequences and to study evolutionary relationships and functional properties \cite{altschul2017sequence}. The Needleman-Wunsch algorithm remains a robust and accurate method for global sequence alignment \cite{NEEDLEMAN1970443}. However, its computational complexity, $O(mn)$, poses significant challenges when processing large-scale data sets \cite{sung2010algorithms}. Worse, when extending the algorithm to multiple sequence alignments the problem becomes NP-complete, requiring exponential time to solve as the problem size increases \cite{wang1994complexity}. This is a serious challenge as it slows down alignment algorithms significantly \cite{liu2010multiple}. In the era of massively parallel computing, new methods need to be developed in an ever-evolving area of biological research.

To address these challenges, this paper introduces a modern high-performance computing approach through a hybrid implementation combining CUDA and MPI. Although both CUDA and MPI have been used independently to solve the problem \cite{gpurmu, 6496050, 8942036}, this hybrid approach utilizes both. The Needleman-Wunsch algorithm is first implemented using CUDA to enable parallel execution of single sequence alignment on GPUs and extended to multiple sequence alignment using MPI to distribute the workload across multiple nodes on a supercomputer. While CUDA handles parallel execution on GPUs, MPI facilitates communication and coordination between nodes, enabling efficient processing of large-scale alignments. 

The combination of CUDA and MPI together has previously been shown to provide a significant acceleration \cite{5429842}. By leveraging CUDA and MPI together, this approach aims to harness the full potential of supercomputing resources to achieve significant performance gains. This paper presents the implementation details and performance evaluation of a Needleman-Wunsch using CUDA and provides an implementation of MPI to be run on a supercomputer. The results demonstrate that such an approach not only accelerates the alignment process, but can see gains with multiple sequence alignments and can address the limitations of traditional CPU-based solutions while not developing the algorithm much further. 

\section{Serial Needleman-Wunsch Algorithm}

The Needleman-Wunsch algorithm is a dynamic programming solution to finding the highest scoring alignment between two sequences by breaking the problem of sequence alignment into the smaller problem of pair matching \cite{NEEDLEMAN1970443}. The algorithm achieves this by constructing a two-dimensional grid of dimensions \( m \times n \) where $m$ and $n$ are the lengths of the sequences being aligned. The steps in full for the algorithm are described as follows:
\begin{enumerate}
    \item Grid Construction
    \item Populate Initial Data
    \item Scoring
    \item Backtracking
\end{enumerate}
The following subsections will cover the procedure for each execution step.

\subsection{Grid Construction}
Accounting for the requirement that each pairwise comparison correlates with its own grid cell, the sequences are aligned in a \( m \times n \) matrix such that the first two elements in the labeling row and column are empty (see Figure 1)

\begin{figure}[h]
    \centering
    \includegraphics{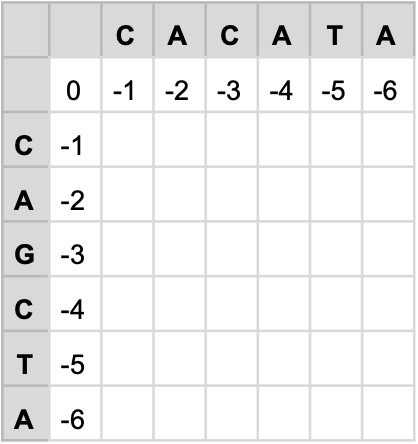}
    \caption{An example two-dimensional alignment grid with primary data as described by the Needleman-Wunsch algorithm}
    \label{fig:Figure 1.}
\end{figure}

\subsection{Populate Initial Data}

Once the \( m \times n \) grid has been initialized and the sequences have been placed along the edges, the top-left cell is populated with a score of \( 0 \), as it does not represent any comparison. The scores for the first row and column are then calculated under the assumption that they represent gaps, resulting in a decreasing sequence along both the row and column. This pattern, which decreases by one for each subsequent cell, establishes the baseline scoring for the alignment grid. For consistency, the algorithm always begins with this configuration, as illustrated in Figure 1, ensuring uniform initialization for all alignments.

\subsection{Scoring}

Scoring methods are largely determined by specific implementations, but generally fall into two main categories: matrix similarity and gap penalty. For this implementation, a gap penalty of \(-1\) was selected due to considerations that will be discussed in the following section. Each cell at location \((i, j)\) is scored based on the formula:  
\begin{align}
\text{score}(i, j) = \text{self} + \max[&\text{score}(i-1, j), \\
&\text{score}(i, j-1), \nonumber \\ &\text{score}(i-1, j-1)] \nonumber
\end{align}
where \(\text{self}\) is the score derived from the match or mismatch of the specific cell's pairing. In this implementation, a match contributes a score of \(+1\), while a mismatch contributes \(-1\). 

Figure 2 illustrates the outcome of this scoring algorithm when applied to the initialized grid shown in Figure 1.

\begin{figure}[h]
    \centering
    \includegraphics{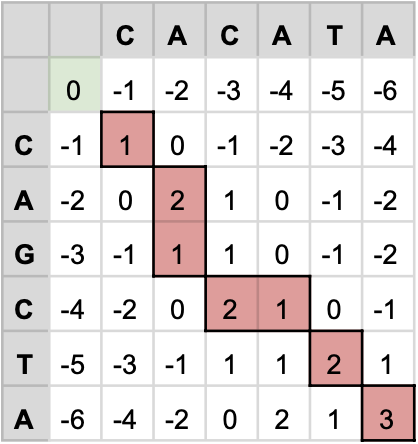}
    \caption{A continuation of the alignment shown in Figure 1 after applying the scoring algorithm.}
    \label{fig:Figure 2.}
\end{figure}

\subsection{Backtracking}
Once the scoring matrix is fully populated, the algorithm determines an optimal alignment by initiating a backtracking process from the bottom-right cell of the matrix. The backtracking follows the path dictated by the \(\max()\) argument from the scoring step, with the following possible directions:

\begin{itemize}
 \item \textbf{Vertical $\mathbf{(i-1,j)}$:} represents a gap in the row sequence
    \item \textbf{Horizontal $\mathbf{(i,j-1)}$:} represents a gap in the column sequence.  
    \item \textbf{Diagonal $\mathbf{(i-1,j-1)}$:} represents either a match or a mismatch with no gap
\end{itemize}

If multiple paths have equal scores during backtracking, all such paths are considered valid alignments. At each branching point, the backtracking process explores all possible options, resulting in multiple potential alignments where ties occur.

\section{Parallel Implementation}

The Needleman-Wunsch algorithm demonstrates high parallelization due to its matrix-based computations, which involve minimal inter-thread dependencies. \cite{8942036,7346733}. The improvements to the algorithm's traditional iterative implementation include using NVIDIA's CUDA library to enable highly parallel GPU calculation of grid cells through partially synchronized threads and reduced computation through the below implementation of the backtrack step of the algorithm. 

Furthermore, to expand beyond the original scope of the Needleman-Wunsch algorithm, MPI is utilized to manage several alignments and perform a heuristic-based center star multiple alignment algorithm due to the continuation of dynamic programming to be an NP-complete problem with exponential complexity \cite{wang1994complexity,ZOU2012322, princeton_msa_lecture}. 

\subsection{Single Alignment}
The original implementation of the Needleman-Wunsch algorithm was designed for single sequence alignment, however, it was developed and published before the popularity of parallel computing and thus implies an extreme iterative solution by calculating each cell one at a time, making three comparisons each time \cite{NEEDLEMAN1970443}. 

In recent times, performance has increased by previously unprecedented margins \cite{Mohamed2020}. With the development of the CUDA suite, we have gained access to parallel code execution across many threads all running in massive parallel schemes due to the specialization of GPU cards \cite{4626815}. 

The kernel implementation presented here is distinct from other approaches by assigning a single thread to each individual cell in the grid, rather than to an entire row, column, or diagonal strip. This design leverages the fact that the computation for a cell depends only on three neighboring cells from the previous step. Threads are manually synchronized with a delay mechanism, allowing calculations to begin as soon as the three dependencies are resolved, rather than waiting for an entire row to complete.

Synchronization is achieved by maintaining a separate, zero-initialized grid that records the backtracking direction for each cell. This direction is encoded as 1 for diagonal, 2 for vertical, or 3 for horizontal. Each thread enters a waiting loop, periodically checking whether all three dependency cells have non-zero values, indicating that their scores have been calculated. Once the dependencies are satisfied, the thread computes the score, updates the backtrack grid with the appropriate direction, and populates the respective cells in both the score and backtrack grids. This approach ensures efficient parallelism by minimizing idle time for threads and optimizing computation flow.

A key feature of the synchronization method in this implementation is the use of a "spinning" loop, which waits for the required dependency cells to complete their computations. By employing this loop instead of relying on a CUDA system call, the need for all threads to wait at a synchronization barrier is eliminated. Instead, each thread begins execution immediately and remains active, only pausing momentarily until the dependent computations from the previous iteration are resolved, ensuring minimal idle time and improved efficiency.

The exact implementation using a CUDA kernel can be seen in Code 1.

\begin{lstlisting}[language=C, caption=CUDA Kernel of Needleman-Wunsch, label=code1, 
basicstyle=\ttfamily\scriptsize, breaklines=true]
__global__ void NW_kernel(int* d_data, int* d_dataComputed, const char* sequence_a, const char* sequence_b){
    // get index
    for( int index = blockIdx.x * blockDim.x + threadIdx.x; index < rows*cols; index += blockDim.x*gridDim.x ){
        int y = floor(index/(double)rows);
        int x = index % cols;
        if((y == 0 || x == 0) || (y > rows || x > cols)) continue;


        int top_index = index - cols;
        int left_index = index-1;
        int diagonal_index = top_index-1;

        while( !d_dataComputed[top_index] || !d_dataComputed[left_index] || !d_dataComputed[diagonal_index] ) { continue; } // wait for values to be filled in d_data

        int top_score = d_data[top_index] * GAP;
        int left_score = d_data[left_index] * GAP;
        int diagonal_score = sequence_a[x] == sequence_b[x] ? d_data[diagonal_index] * MATCH : d_data[diagonal_index] * MISMATCH;

        d_dataComputed[index] = 1;
    }

}
\end{lstlisting}

\subsection{Multiple Alignment}
A more advanced approach is to extend the single alignment implementation to multiple sequence alignment (MSA). Utilizing CUDA and parallel computation, one can align tens or even hundreds of sequences simultaneously, greatly improving efficiency. MPI can facilitate the distribution of the workload between multiple nodes in a cluster, enhancing scalability.

Since MSA is NP-complete, parallelizing the computation with CUDA and MPI enables us to handle larger datasets and achieve faster alignment times. This approach leverages the computational power of GPUs and the scalability of distributed computing, making it suitable for analyzing complex biological sequences on modern high-performance computing systems.

There are several multiple sequence alignment algorithms. For simplicity, the simple center star multiple sequence alignment algorithm was chosen \cite{ZOU2012322, princeton_msa_lecture}. This algorithm utilizes a heuristic to find the sequence that is most similar to the rest of the sequences (center sequence) through a scoring matrix. A simple scoring method that assigns a positive score for matches and a negative score for mismatches is used.

After choosing a center sequence, align all pairwise sequences with the center. Next, iteratively merge the alignments using the aligned center sequence as a reference. For the implementation presented, only the scoring matrix computation and alignment stages are evaluated, as these are the most computationally intensive and effectively leverage both CUDA and MPI for efficient parallel processing.

In MPI, the total number of alignments is divided by the number of ranks to assign each rank a fair share of the computation. The data is then sent to each rank and a CUDA kernel is launched to align the sequences using the single-sequence alignment implementation presented previously. After completion, the data is gathered back in the main process for further processing and merging. 
If there are $n$ sequences of length $k$ then one must compute:
\begin{align}
    \frac{n(n-1)}{2} = \frac{n^2-n}{2} = O(n^2)
\end{align}
Thus, the total running time is $O(n^2k^2)$. Although this does not seem ideal, there may be significant speedups by utilizing MPI and CUDA together to run the alignment process across several GPU nodes in a distributed system environment.

\section{Analysis}
The analysis focuses on evaluating the CUDA aspect of the implementation, specifically examining the scaling performance of single sequence alignment. This limitation is imposed by the current hardware constraints, which prevent testing at larger scales or incorporating distributed computation. Nevertheless, the observed results provide meaningful insights into the efficiency and performance improvements achieved through GPU-based parallelization.

\subsection{Strong Scaling}
In strong scaling experiments, the problem size is kept constant and progressively increases the number of threads allocated to the program. To prove this, the sequences of two gene sequences (COL1A1 and THAP11) from \textit{M. musculus} and \textit{H. sapiens} were chosen for alignment.

As shown in Figure 3 (COL1A1) and Figure 4 (THAP11), the runtime decreased significantly as thread counts increased, demonstrating the effectiveness of this parallel implementation approach compared to the serial version.

\begin{figure}[h]
    \centering
    \includegraphics[width=\columnwidth]{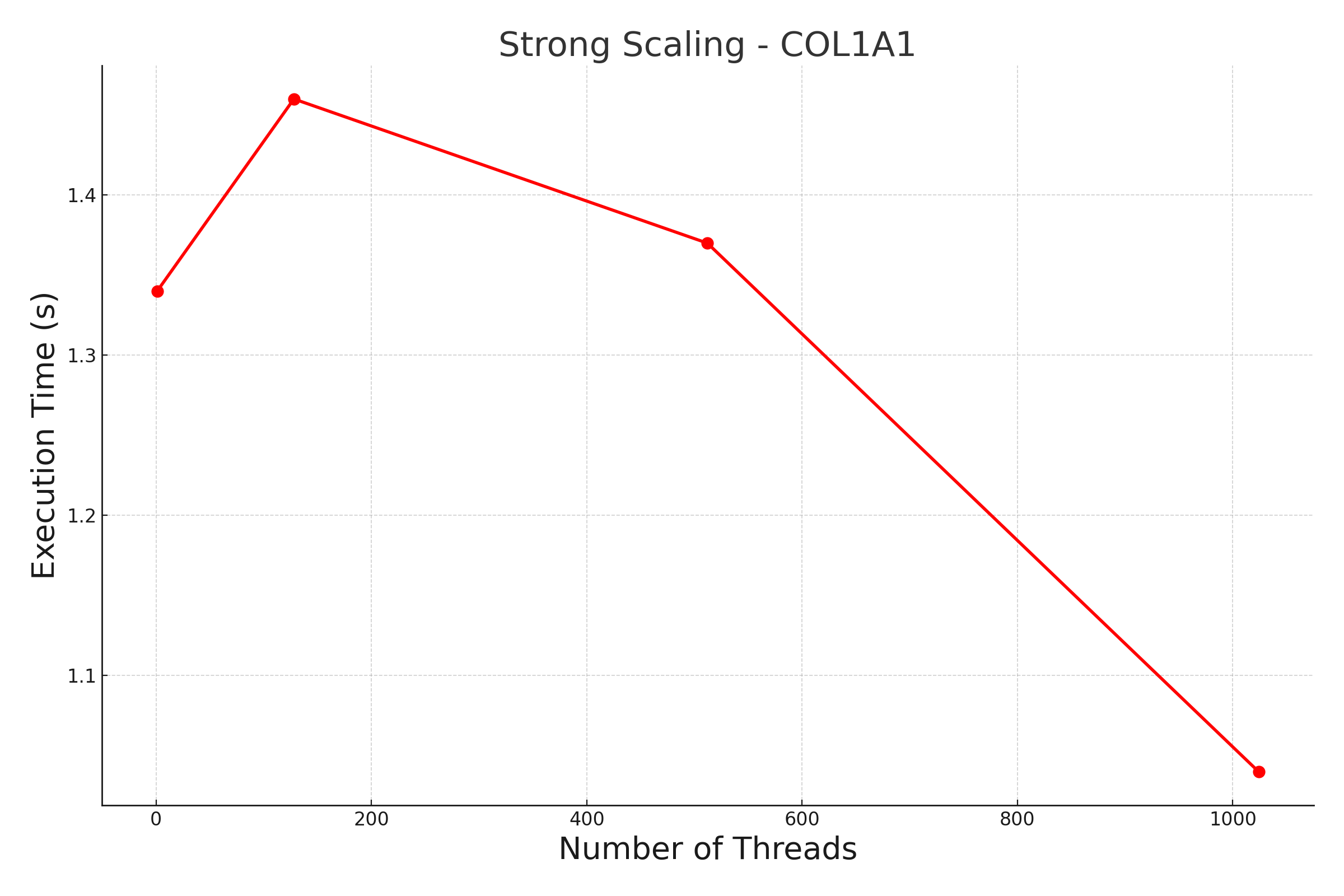}
    \caption{Strong scaling comparison of a COL1A1 sequence alignment between mouse and human}
    \label{fig:Figure 3.}
\end{figure}

\begin{figure}[h]
    \centering
    \includegraphics[width=\columnwidth]{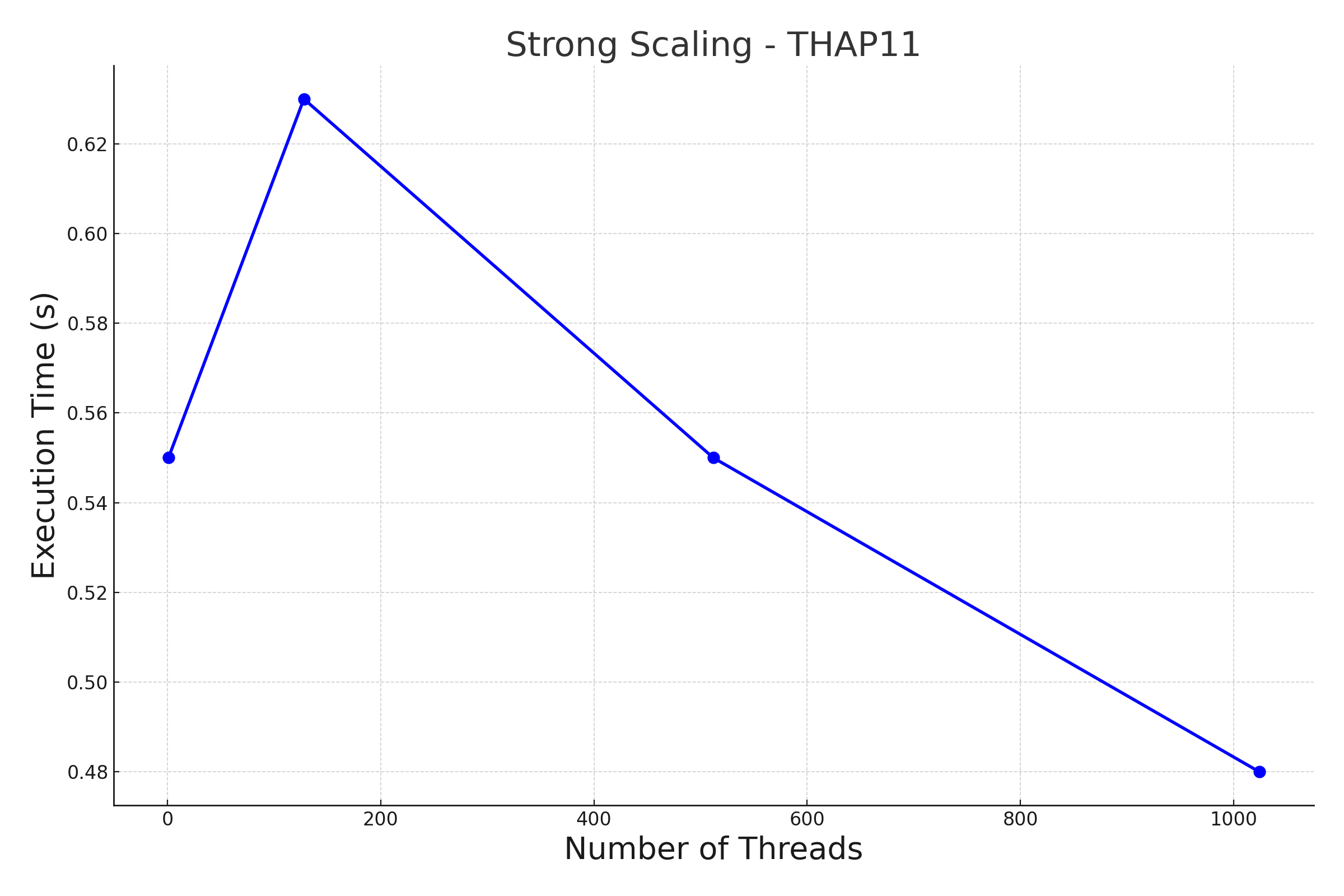}
    \caption{Strong scaling comparison of a THAP11 sequence alignment between mouse and human}
    \label{fig:Figure 4.}
\end{figure}

In the serial code, each cell is calculated sequentially, one at a time, with dependencies resolved row by row. This sequential approach results in significant idle time, as computations cannot start on cells that rely on incomplete dependencies. In contrast, the parallel implementation presented assigns one thread per cell rather than per row, enabling much finer-grained parallelism. Once the dependencies of a cell are satisfied, the corresponding thread can immediately perform its calculation and be reallocated to subsequent cells.

The key observation from the graphs is the consistent reduction in runtime as the number of threads increases. This continual improvement is largely due to the design of the CUDA kernel provided, where:

\begin{enumerate}
    \item Threads operate on individual cells, avoiding bottlenecks caused by entire rows waiting for dependencies.
    \item Freed threads are quickly reallocated to other ready-to-process cells, ensuring minimal thread idle time.
\end{enumerate}

This design achieves near zero wait time for thread reallocation, resulting in highly efficient usage of available computational resources. Even at high thread counts (e.g., 1024 threads), the runtime continues to improve, suggesting that the workload is efficiently distributed across all threads, avoiding overhead from load imbalance.

The variation in performance at lower thread counts, particularly for 128 threads in both graphs, could be attributed to kernel launch overhead or thread synchronization delays. However, as thread counts increase further, these overheads are amortized, and the implementation scales effectively.

\subsection{Weak Scaling}
The weak scaling experiment evaluates the performance of the implementation as the problem size increases proportionally to the number of available threads. This analysis serves to assess how well the algorithm maintains efficiency under growing computational workloads while additional resources are introduced. The results, presented in Figure 5, compare the execution times of the serial and parallel implementations.

\begin{figure}
    \centering
    \includegraphics[width=\columnwidth]{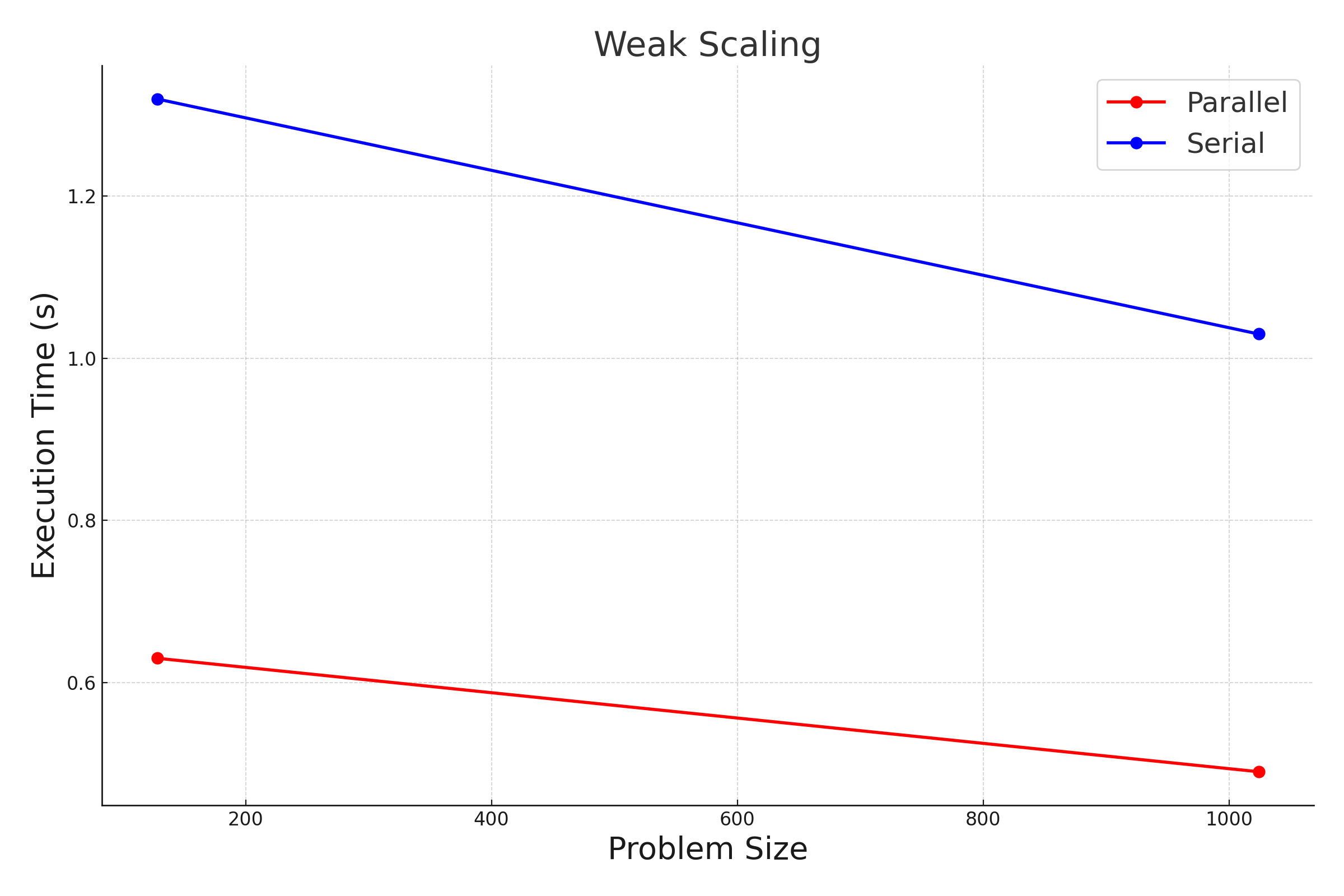}
    \caption{Weak scaling comparison of execution times as problem size increases}
    \label{fig:Figure 5.}
\end{figure}

In the serial implementation, execution time exhibits a linear decline as the problem size scales upward. However, the rate of decline (slope) is noticeably steeper than that of the parallel implementation. This behavior, while initially counterintuitive, suggests a potential inefficiency in the serial algorithm's handling of increasing problem sizes. As workloads grow, the serial code may benefit disproportionately from larger problem sizes due to factors such as improved cache utilization, reduced context switching, or streamlined memory access patterns. However, this steep decline also highlights an inherent limitation of the serial code: while it is beneficial under weak scaling conditions, the sequential nature of its design prevents it from achieving further meaningful acceleration.

In contrast, the parallel implementation demonstrates a more gradual reduction in execution time as the problem size increases. This result reflects a key characteristic of the parallel design: while the workload is efficiently distributed across threads, the additional overheads of parallel execution-such as kernel launch latency, thread synchronization, and memory management persist at larger scales. As a result, the slope of the execution time reduction is less pronounced. Although the parallel implementation remains significantly faster than its serial counterpart, the weaker slope may indicate areas for further optimization, particularly in minimizing thread synchronization and memory access overheads as workloads grow.

The disparity in slopes between the serial and parallel implementations is a critical observation. It underscores both the strengths and potential limitations of the parallel approach. While the parallel implementation maintains excellent performance and efficiency across larger workloads, the diminishing returns observed at scale suggest that improvements to the kernel design, thread allocation strategy, or memory optimization may yield further gains. Conversely, the rapid decline of execution time in the serial implementation, while seemingly advantageous, reflects a limitation of sequential computation that cannot be overcome for larger-scale tasks.

In summary, the weak scaling analysis highlights the superior scalability and overall performance of the parallel implementation compared to its serial counterpart. However, the flatter slope of the parallel execution time indicates opportunities for further refinement, particularly to address overheads that become increasingly prominent at larger problem sizes. This observation provides a foundation for future work aimed at enhancing the efficiency and scalability of the parallel implementation. 

\section{Conclusion}

In this paper, a stable, parallelized implementation of the Needleman-Wunsch algorithm designed to run on CUDA-enabled GPUs was presented. The implementation demonstrated substantial improvements in computational efficiency for deterministic global sequence alignments. These results show that future multiple sequence alignment work may show significant performance gains as well. 

Given that the dynamic programming approach to multiple sequence alignment is an NP-complete problem with exponential time complexity \cite{wang1994complexity}, this paper provides a foundation for future improvements. Further work is needed to advance the development of multiple sequence alignment algorithms and to conduct a more comprehensive analysis of their results. Enhancing the efficiency of the implementation remains a critical area of focus, particularly in minimizing the likelihood of missing an optimal alignment.

Furthermore, advancements in CUDA-MPI integration offer significant potential to further accelerate performance and scalability, enabling the alignment of larger datasets within distributed computing environments. These improvements will contribute to addressing the computational challenges associated with large-scale sequence alignment such as entire genome alignment.

\bibliographystyle{unsrt}
\bibliography{example}

\begin{thebibliography}{10}

\bibitem{altschul2017sequence}
Stephen~F. Altschul and Mihai Pop.
\newblock Sequence alignment.
\newblock In Kenneth~H. Rosen, Douglas~R. Shier, and William Goddard, editors,
  {\em Handbook of Discrete and Combinatorial Mathematics}, chapter 20.1. CRC
  Press/Taylor \& Francis, Boca Raton, FL, 2nd edition, Nov 2017.

\bibitem{NEEDLEMAN1970443}
Saul~B. Needleman and Christian~D. Wunsch.
\newblock A general method applicable to the search for similarities in the
  amino acid sequence of two proteins.
\newblock {\em Journal of Molecular Biology}, 48(3):443--453, 1970.

\bibitem{sung2010algorithms}
Wing-Kin Sung.
\newblock {\em Algorithms in Bioinformatics: A Practical Introduction}.
\newblock Chapman \& Hall/CRC Press, Boca Raton, 2010.

\bibitem{wang1994complexity}
Lusheng Wang and Tao Jiang.
\newblock On the complexity of multiple sequence alignment.
\newblock {\em Journal of Computational Biology}, 1(4):337--348, Winter 1994.

\bibitem{liu2010multiple}
Kevin Liu, C.~Randal Linder, and Tandy Warnow.
\newblock Multiple sequence alignment: a major challenge to large-scale
  phylogenetics.
\newblock {\em PLOS Currents Tree of Life}, Nov 2010.

\bibitem{gpurmu}
Chun-Yuan Lin, Yu-Shiang Lin, Jiayi Zhou, and Chuan Tang.
\newblock Gpu-remusic: Efficient constrained multiple sequence alignment
  algorithm on graphics processing units.
\newblock 01 2011.

\bibitem{6496050}
Da~Li and Michela Becchi.
\newblock Abstract: Multiple pairwise sequence alignments with the
  needleman-wunsch algorithm on gpu.
\newblock In {\em 2012 SC Companion: High Performance Computing, Networking
  Storage and Analysis}, pages 1471--1472, 2012.

\bibitem{8942036}
Veska Gancheva and Ivaylo Georgiev.
\newblock Multithreaded parallel sequence alignment based on needleman-wunsch
  algorithm.
\newblock In {\em 2019 IEEE 19th International Conference on Bioinformatics and
  Bioengineering (BIBE)}, pages 165--169, 2019.

\bibitem{5429842}
N.~P. Karunadasa and D.~N. Ranasinghe.
\newblock Accelerating high performance applications with cuda and mpi.
\newblock In {\em 2009 International Conference on Industrial and Information
  Systems (ICIIS)}, pages 331--336, 2009.

\bibitem{7346733}
Anuj Chaudhary, Deepkumar Kagathara, and Vibha Patel.
\newblock A gpu based implementation of needleman-wunsch algorithm using
  skewing transformation.
\newblock In {\em 2015 Eighth International Conference on Contemporary
  Computing (IC3)}, pages 498--502, 2015.

\bibitem{ZOU2012322}
Quan Zou, Xiao Shan, and Yi~Jiang.
\newblock A novel center star multiple sequence alignment algorithm based on
  affine gap penalty and k-band.
\newblock {\em Physics Procedia}, 33:322--327, 2012.
\newblock 2012 International Conference on Medical Physics and Biomedical
  Engineering (ICMPBE2012).

\bibitem{princeton_msa_lecture}
Mona Singh.
\newblock Lecture 4: Multiple sequence alignments i, 2000.
\newblock Lecture notes, COS 551: Introduction to Computational Molecular
  Biology, Princeton University.

\bibitem{Mohamed2020}
Khaled~Salah Mohamed.
\newblock {\em Parallel Computing: OpenMP, MPI, and CUDA}, pages 63--93.
\newblock Springer International Publishing, Cham, 2020.

\bibitem{4626815}
Michael Garland, Scott Le~Grand, John Nickolls, Joshua Anderson, Jim Hardwick,
  Scott Morton, Everett Phillips, Yao Zhang, and Vasily Volkov.
\newblock Parallel computing experiences with cuda.
\newblock {\em IEEE Micro}, 28(4):13--27, 2008.

\end{thebibliography}

\end{document}